\documentstyle[preprint,version2,aps]{revtex}

\begin{document}
\draft
\begin{title}
X-ray Resonant Scattering Studies of Orbital and Charge Ordering
in Pr$_{1-x}$Ca$_x$MnO$_3$
\end{title}

\author{
M. v. Zimmermann$^1$, C.S. Nelson$^1$, J. P. Hill$^1$, Doon Gibbs$^1$,
M. Blume$^1$,
D. Casa$^2$, B. Keimer$^2$, Y. Murakami$^3$, C.-C. Kao$^4$,
C. Venkataraman$^5$, T. Gog$^5$
Y. Tomioka$^6$ and Y. Tokura$^{6.7}$}

\begin{instit}
$^1$ Department of Physics, Brookhaven National Laboratory
Upton, NY 11973-5000

$^2$ Department of Physics, Princeton University
Princeton, NJ 08544
and
Max-Planck-Institut fur Festoperforschung
70569 Stuttgart, Germany

$^3$ Photon Factory, Institute of Materials Structure Science
High Energy Accelerator Research Organization
Tsukuba 305-0801, Japan

$^4$ National Synchrotron Light Source, Brookhaven National Laboratory
Upton, NY  11973-5000

$^5$ CMC-CAT, Advanced Photon Source, Argonne National Laboratory
Argonne, IL 60439

$^6$ Joint Research Center for Atom Technology (JRCAT)
Tsukuba 305-0033, Japan

$^7$ Department of Applied Physics, University of Tokyo
Tokyo 113-0033, Japan
\end{instit}

\begin{abstract}

We present the results of a systematic x-ray scattering study of
the charge and orbital ordering in the manganite series
Pr$_{1-x}$Ca$_x$MnO$_3$ with $x$=0.25, 0.4 and 0.5. The
temperature dependence of the scattering at the charge and orbital
wavevectors, and of the lattice constants, was characterized
throughout the ordered phase of each sample. It was found that the
charge and orbital order wavevectors are commensurate with the
lattice, in striking contrast to the results of earlier electron
diffraction studies of samples with $x$=0.5. High
momentum-transfer resolution studies of the x=0.4 and 0.5 samples
further revealed that while long-range charge order is present,
long-range orbital order is never established.  Above the
charge/orbital ordering temperature T$_o$, the charge order
fluctuations are more highly correlated than the orbital
fluctuations.  This suggests that charge order drives orbital
order in these samples. In addition, a longitudinal modulation of
the lattice with the same periodicity as the charge and orbital
ordering was discovered in the x=0.4 and 0.5 samples. For x=0.25,
only long-range orbital order was observed with no indication of
charge ordering, nor of an additional lattice modulation.  We also
report the results of a preliminary investigation of the loss of
charge and orbital ordering in the x=0.4 sample by application of
a magnetic field. Finally, the polarization and azimuthal
dependence of the charge and orbital ordering in these compounds
is characterized both in the resonant and nonresonant limits, and
compared with the predictions of current theories. The results are
qualitatively consistent with both cluster and LDA+U calculations
of the electronic structure.

\end{abstract}

\newpage

\section{Introduction}

Interest in the origins of high temperature superconductivity and
colossal magnetoresistance in the transition metal oxides has driven
much of the activity currently at the center of condensed matter
physics. An important aspect of these strongly correlated electron
systems is that no single degree of freedom dominates their response.
Rather, the ground state properties are thought to reflect a balance
among several correlated interactions, including orbital and charge
ordering, magnetism, and coupling to the lattice.

The perovskite manganites provide an especially illuminating
example of the interplay among these interactions, since in these
materials the balance may be conveniently altered, for example by
doping or through an applied magnetic field. As a result, much
work has been done to understand their magnetic ground states and
lattice distortions, dating back to the seminal experiments of
Wollan and Koehler \cite{Wollan55}.  Less is known about the roles
of charge and orbital order in these materials.  The classic work
of Goodenough \cite{Goodenough55}  has nevertheless served as a
guide to their ordered arrangements, as supplemented, for example,
by detailed measurements of the crystal structure and of the
temperature dependence of the lattice constants (see references
\cite{Jirak85,Radaelli97}, for example).

This situation has changed during the last two years following the
detection of orbital and charge order by resonant x-ray scattering
techniques
\cite{Murakami214,Murakami113,Ishihara-PRL-1998,Fabrizio98,Elfimov99,Endoh99,Fulde}.
Specifically, it has been found that the sensitivity of x-ray
scattering to these structures can be significantly enhanced by
tuning the incident x-ray energy to the transition metal
K-absorption edge. Thus, it appears possible to characterize the
orbital and charge ordering on a microscopic scale, and to study
their response to changes of temperature or to an applied magnetic
field. Insofar as we are aware, resonant x-ray scattering studies
of these materials have now been extended to include 
La$_{0.5}$Sr$_{1.5}$MnO$_4$  \cite{Murakami214}, LaMnO$_3$
\cite{Murakami113}, La$_{1-x}$Sr$_{x}$MnO$_3$
\cite{Endoh99,Wochner_unpub}, Pr$_{1-x}$Ca$_x$MnO$_3$
\cite{Zimmermann99}, V$_2$O$_3$ \cite{Paolosini99}, YTiO$_3$
\cite{Nakao_unpub}, LaTiO$_3$\cite{Keimer2000},
LaSr$_2$Mn$_2$O$_7$ \cite{Wakabayashi_unpub},
DyB$_2$C$_2$\cite{Tanaka99,Hirota2000}, and
Nd$_{0.5}$Sr$_{0.5}$MnO$_3$\cite{Nakamura99,Hatton99},--and this
list continues to grow. There is, in addition, an ongoing
discussion of whether it is more appropriate to treat the resonant
cross-section within an extended, band structure description of
the electronic structure, or instead with a more localized, atomic
description \cite{Elfimov99,Benfatto99,Ishihara98,Fulde}. A
related question concerns how to write the resonant cross-section
explicitly in terms of the order parameters for orbital and charge
ordering.

In this paper, we present x-ray scattering studies of
Pr$_{1-x}$Ca$_x$MnO$_3$ with x=0.25, 0.4 and 0.5. Detailed studies
have been made of the temperature dependence of the orbital and
charge order scattering of all three samples, including
characterization of the intensities, wavevectors, correlation
lengths and lattice constants. Below a doping-dependent ordering
temperature T$_o$, it is found that the charge and/or orbital
order wavevectors are commensurate with the lattice at all
temperatures. This contrasts with the results of electron
diffraction studies of samples with x=0.5, where a significant
variation of the charge order wavevector was reported near
T$_o$\cite{Chen99,Jirak2}. Surprisingly, our high momentum-transfer
resolution studies reveal that long-range orbital order is never
established in the x=0.4 and 0.5 samples, although long-range
charge order is observed in both.  Further, for temperatures above
T$_o$, the charge order fluctuations are longer ranged than the
orbital fluctuations, suggesting that the charge ordering drives
the orbital ordering in these systems. Recent Landau theories of
the phase transition are consistent with this picture \cite{Zhang99}. In
contrast, for x=0.25 we observe long-range orbital order, with no
indication of any charge ordering.  We have also monitored the
destruction of charge and orbital ordering after the application
of a magnetic field in the x = 0.4 sample.  A similar
phenomenology is found for increasing magnetic field as occurs for
increasing temperature.

Finally, detailed measurements of the polarization and azimuthal
dependence of the charge and orbital ordering have been carried
out in both the resonant and nonresonant limits. In the $\sigma
\rightarrow \pi$ channel at the orbital wavevector of all three
samples, we find that the resonant cross-section is qualitatively
consistent with the results obtained earlier for LaMnO$_3$
\cite{Murakami113} and with the predictions of both the localized
and band-structure descriptions of the electronic structure.
Likewise, we have found that the resonant scattering at the charge
order wavevector is consistent with earlier results obtained for
La$_{1.5}$Sr$_{0.5}$MnO$_4$ \cite{Murakami214}.  We have, in
addition, discovered scattering in the $\sigma \rightarrow \sigma$
channel at the charge and orbital wavevectors of the x=0.4 and 0.5
samples. On the basis of its polarization and Q-dependence, we
have deduced that it originates from a longitudinal lattice
modulation. Earlier studies of La$_{0.5}$Ca$_{0.5}$MnO$_3$, had
previously found a transverse modulation
\cite{Radaelli97,Radaelli99,Wang_sub,Chen90}, and a similar
modulation has been assumed in Pr$_{1-x}$Ca$_x$MnO$_3$.  A summary
of the present work was published earlier \cite{Zimmermann99}.

The organization of this paper is as follows.  The experimental
set-up is described immediately below, followed by a brief
description of charge and orbital ordering in
Pr$_{1-x}$Ca$_x$MnO$_3$.  A simple model of the resonant orbital
cross-section is given in Section IV. Our main results and
discussion follow in Section V, which is divided into A.
Diffraction Pattern, B. Resonant Scattering, C. Temperature
Dependence, and D. Magnetic Field Dependence. A brief summary is
given at the end.

\section{Experimental}

The single crystals used in the present experiments were grown by
float zone techniques at JRCAT. (0,1,0) surfaces
were cut from cylinders of radius ~3 mm, and polished with fine
emery paper and diamond paste.  The mosaic widths of the samples
as characterized at the (0,2,0) bulk Bragg reflections (in
orthorhombic notation) were 0.1, 0.25, and 0.25$^{\circ}$, (FWHM),
for the x=0.25 0.4 and 0.5 samples, respectively. These values
varied by small amounts as the beam was moved across each sample
surface, reflecting its mosaic distribution. The growth techniques
and basic transport properties of these crystals have been
described in detail elsewhere \cite{Okimoto98,Tomioka96,Tokura}.

Most of the x-ray scattering experiments were carried out at the
National Synchrotron Light Source on Beamlines X22C, X22B and X21.
X22C is equipped with a bent, toroidal focussing mirror and a
Ge(111) double crystal monochromator arranged in a vertical
scattering geometry. This gives an incident linear polarization of
95\% ($\sigma$) and an incident energy resolution of about 5 eV at
the Mn K edge (6.545 keV). Three different detector configurations
were used. Low momentum-transfer resolution scans employed slits
before the detector, and provided a longitudinal resolution of
0.0021$\AA^{-1}$(HWHM) at the (010) reflections of each sample.
High-resolution scans employed a standard Ge(111) crystal, and
gave a longitudinal resolution of $4.5 x 10^{-4}$$\AA^{-1}$(HWHM)
at the respective (010) reflections . The third configuration
provided linear polarization analysis of the scattered beam via
rotation of a Cu(220) crystal about the scattered beam direction
\cite{Gibbs88}.  It gave longitudinal resolutions of 0.0069
$\AA^{-1}$ and 0.0052$\AA^{-1}$ (HWHM) in the $\sigma \rightarrow
\sigma$ and $\sigma \rightarrow \pi$ geometries, respectively. For
an incident photon energy set at the Mn K absorption edge, the
Cu(220) scattering angle is 95.6$^{\circ}$. This leads to a 5-10\%
uncertainty in the polarization-dependent intensities due to
incomplete suppression of the unselected component of the
scattered beam, and the small $\pi$-component of the incident
beam.

Magnetic field experiments were performed at NSLS X22B, which
supports a bent, toroidal mirror and a single crystal Ge(111)
analyzer-monochromator combination. As a result of mechanical
restraints, the incident photon energy could not reach the Mn K
edge. These experiments were, therefore, carried out in the
nonresonant limit, with an incident photon energy of 8 keV. The
sample was mounted in a 13 T superconducting magnet oriented in a
horizontal scattering geometry. In addition, two series of
experiments were performed on NSLS Wiggler beamline X21, which was
equipped with a 4-bounce Si(220) monochromator and a focussing
mirror, leading to extremely good incident energy resolution of
0.25 eV. One set of experiments was carried out on undulator
beamline 9ID at the Advanced Photon Source.  The optics for 9ID
was comprised of a double crystal Si(111) monochromator and a flat
harmonic rejection mirror.

\section{ Proposed Orbital and Charge Ordered Structures}

At room temperature, the crystal structure of Pr$_{1-x}$Ca$_x$MnO$_3$ is
orthorhombic (Pbmn), as illustrated in Figure 1. Characteristic of the
perovskite manganites, each Mn atom lies at the center of the octahedron
defined by the oxygen atoms at the corners.  Single layers of Pr atoms
lie between the layers of octahedra. Depending on the temperature, there
may be distortions of the octahedra and tilts as is
also illustrated in Fig. 1.  The solid line in the figure outlines the
orthorhombic unit cell.

A schematic phase diagram for Pr$_{1-x}$Ca$_x$MnO$_3$ versus Ca
concentration and temperature \cite{Jirak85,Tomioka96} is shown in Figure
2. For small x $(0.15 \leq x \leq 0.3)$ and at low temperatures,
Pr$_{1-x}$Ca$_x$MnO$_3$ is a ferromagnetic insulator, and is
believed to exhibit an orbitally ordered ground state analogous to
that observed in LaMnO$_3$. The electronic configuration of the
Mn$^{3+}$ (d$^4$) ions is ($t^3_{2g}$, $e_g^1$) with the $t_{2g}$
electrons localized at the Mn sites. The $e_g$ electrons are
hybridized with the oxygen 2p orbitals, and believed to
participate in a cooperative Jahn-Teller distortion of the MnO$_6$
octahedra. This leads to a ($3x^2-r^2$)-($3y^2-r^2$)-type of
orbital order of the $e_g$ electrons in the ab plane with the
oxygens displaced along the direction of extension of the $e_g$
orbitals. A schematic illustration of this orbitally ordered state
for x=0.25 is shown in Figure 3a\cite{Jirak85}, with the orbital
unit cell marked by the solid line. The excess Mn$^{4+}$ ions in
this material are believed to be disordered.  It is noteworthy
that the orbital period is twice that of the fundamental Mn
spacing, so that orbital scattering appears at structurally
forbidden reflections. In orthorhombic notation, for which the
fundamental Bragg peaks occur at (0,2k,0), the orbital scattering
then occurs at (0,k,0).

Recently, the possibility of the existence of both charge and
orbital ordering at x=0.25 has been suggested by various
theoretical approaches\cite{Mizokawa2000,Dagotto}. Mizokawa, et
al.\cite{Mizokawa2000} studied a related material,
La$_{1-x}$Sr$_x$MnO$_3$, and found an ordered arrangement of
($3x^2-r^2$)-($3y^2-r^2$)and ($3z^2-r^2$) type orbitals
surrounding the Mn$^{4+}$ sites at x=0.25.  While this structure
is inconsistent with the magnetic structure in
Pr$_{0.75}$Ca$_{0.25}$MnO$_3$, it first raised the possibility of
structures other than those proposed by Jirak, et al.
\cite{Jirak85}. As discussed below, however, we have found no evidence
for this type of charge ordering.

For Ca concentrations 0.3 $\leq$ x  $\leq$0.7,
Pr$_{1-x}$Ca$_x$MnO$_3$ becomes an antiferromagnetic insulator at
low temperatures (see Figure 2), and exhibits colossal
magnetoresistance in applied magnetic fields, with the
metal-insulator transition occurring between 5 and 8
T\cite{Tomioka96}.  These effects result from charge ordering
among the Mn$^{3+}$ and Mn$^{4+}$ ions, which occurs in addition
to orbital ordering. The large conductivity is enabled through the
hopping of $e_g$ electrons among Mn sites. The fraction of Mn ions
in the Mn$^{4+}$ state is determined largely by the concentration
of Ca ions. Thus, by varying the Ca concentration, it is possible
to alter the balance between charge and orbital ordering. The
proposed ground state\cite{Jirak85} for both the x=0.4 and x=0.5
concentrations is shown in Figure 3b. The small filled circles
represent the Mn$^{4+}$ ions, with one fewer electron than is
localized at the Mn$^{3+}$ sites. The solid line indicates the
unit cell for orbital ordering, while the dashed line gives that
for charge ordering. It is interesting that the proposed
structures for x=0.4 and 0.5 are identical\cite{Jirak85} and
commensurate with the lattice, independent of concentration.
Clearly, at least for x = 0.4, this picture cannot be strictly
correct.  Jirak {\em et al.} proposed that the extra electrons
present at x=0.4 could be accommodated in such a structure by a
partial occupancy of the $3z^2- r^2$ orbitals of the nominal
Mn$^{4+}$ sites. Other possibilities include small Mn$^{3+}$ rich
regions, higher order structures, or small regions of orbital
disorder. As discussed below, our data reveal that, in fact, the
orbital order is not long-ranged in either of these compounds,
although the charge order is. In the orthorhombic notation, the
charge order reflections occur at (0,2k+1,0) and the orbital order
reflections at (0,k+1/2, 0). Note that the orbital period ($=2b$) in the
x=0.4, 0.5 compounds differs from that occurring in samples with
x$<$0.3 ($=b$), as a result of the presence of charge ordering.

The magnetic structure of these compounds at low doping (0.15
$\leq$ x $\leq$ 0.3) is ferromagnetic with T$_C \approx$140 K.
Compounds with higher doping (0.3 $\leq$ x $\leq$ 0.75) are
CE-type antiferromagnets with T$_N$=170 K for x between 0.4 and
0.5\cite{Jirak85}.  The in-plane components of the magnetic
structures are also illustrated in Figure 3.

\section{Resonant Cross-section for Orbital Ordering--A Simple Model}

The present experiments were carried out using x-ray resonant
scattering techniques.  As shown in a series of recent papers
\cite{Murakami214,Murakami113,Ishihara-PRL-1998,Fabrizio98,Elfimov99,Endoh99,Benfatto99,Wochner_unpub,Nakamura99,Ishihara98,Fulde},
the sensitivity of x-ray scattering to orbital ordering in the
transition metal oxides is enhanced when the incident x-ray energy
is tuned near the K-absorption edge. In the resonant process, a
core level electron is promoted to an intermediate excited state,
which subsequently decays. This can lead to new scattering
mechanisms, such as resonant magnetic scattering. In that case,
the excited electron is promoted to a partially occupied orbital
and the sensitivity to the magnetic polarization arises through
the exchange interaction \cite{Blume_book,Hannon88}.  In the
present case, we consider a dipole process involving a Mn 1s to 4p
transition. In the simplest model, it is assumed that the
4p$_{x,y,z}$ states are initially unoccupied, but split in the
orbitally ordered state (see figure 4). This gives rise to a
nonzero resonant scattered intensity at reflections sensitive to
the difference between the two orbitally ordered sublattices. This
model, summarized below, is designed to capture the essence of the
problem, but not the details of the interactions for which a more
sophisticated theory is required. For example, recent LSDA+U
calculations \cite{Elfimov99} suggest that the 4p bands are not
split {\em per se}, but rather experience changes in the weight of
the density of states in the ordered phase. Nevertheless, our
model is conceptually simple, and reproduces most of the
systematics of the data.

For a difference reflection, the resonant scattered intensity may be
written \cite{Murakami113}:
\begin{equation}
I^{res}= \sum_{x,y,z,n=\pm 1} n
\frac{<s|P^{\alpha}|p_m><p_m|P^{\beta}|s>}{\omega - \omega_o
-\delta\omega^n_m + i\Gamma/2} \epsilon^{'\alpha}\epsilon^{\beta},
\end{equation}
where the coordinate system has been chosen so that x and y are
along the direction of extension of the ordered $e_g$ orbitals and
z is perpendicular to the x-y plane (see Figure 4). $|s>$ and
$|p>$ are the wavefunctions of the Mn 1s and 4p orbitals,
respectively.  P$^{\alpha}$ is the $\alpha$ component of the
dipole operator ($\alpha$ = x,y,z). $\omega$ is the incident
photon energy and $\omega_o$ is the energy of the unperturbed
p$_m$ levels.  The incident (final) polarization of the photons is
$\epsilon$($\epsilon^{,}$) and n=$\pm$1 labels the orbital
sublattice. $\Gamma$ is the lifetime of the excited state.  Note,
$<p_m|P^{\alpha}|s> = A\delta_{m\alpha}$, where A is a constant.
As illustrated in the figure, $\delta^n_m=-\Delta$ for n=+1,
m=x,z; $\delta^n_m=2\Delta$ for n=+1, m=y; and so on.

The origin of the splitting $\Delta$ is not specified in our
model.  Two mechanisms (both consistent with the measurements)
have been discussed, and indeed, the discussion has sparked some
controversy
\cite{Murakami214,Murakami113,Ishihara-PRL-1998,Fabrizio98,Elfimov99,Endoh99,Benfatto99,Wochner_unpub,Hatton99,Ishihara98,Fulde}.
One possible origin involves the Coulomb coupling of the Mn 3d and
4p levels, either directly or indirectly through the hybridization
of the Mn(3d)-0(2p) and 0(2p)-Mn(4p) states \cite{Ishihara98}. The latter
effect has the same sign as the direct Coulomb interaction, but is
expected to be smaller\cite{Ishihara98}. In this picture, the
Coulomb coupling raises the 4p$_m$ levels lying parallel to the
direction of extension of the orbital (by 2$\Delta$ in our model)
and lowers those lying perpendicular (by $\Delta$) as shown in
Figure 4. Detailed calculations using atomic orbitals by Ishishara
and Maekawa \cite{Ishihara98} have found qualitative consistency
with all of the known experimental results for
La$_{1-x}$Sr$_{1+x}$MnO$_4$\cite{Murakami214}, LaMnO$_3$
\cite{Murakami113}, Pr$_{1-x}$Ca$_x$MnO$_3$\cite{Zimmermann99},
and La$_{1-x}$Sr$_{x}$MnO$_3$ \cite{Endoh99}. Alternatively, the
motion of the oxygen atoms away from regions of high charge
density through the Jahn-Teller interaction lowers the energy of
the 4p$_m$ levels lying parallel to the direction of the extension
of the orbital, and raises those lying perpendicular. This effect
thus has the opposite sign to that of the 3d-4p Coulomb
interaction discussed above, and in fact these mechanisms compete
with each other. Several groups have argued that the oxygen motion
is the dominant effect leading to resonant scattering at the
orbital wavevector\cite{Elfimov99,Benfatto99,Fulde,Benedetti2000,Takahashi99}.
Calculations of the resonant cross-section based on such
approaches, and utilizing band structure descriptions of the 4p
density of states in LaMnO$_3$ (which show changes in weight,
rather than a simple splitting) also reproduce the main
experimental facts and further, make detailed predictions about
the resonant fine structure measured in x-ray scattering
experiments\cite{Elfimov99,Zimmermann99}. Insofar as we are aware,
the experimental data obtained to date do not distinguish either
theoretical approach conclusively, and this remains an open
question. For the purpose of calculating the resonant
cross-section of our simplified model, however, all that is
required is that $\Delta \neq 0$.

We stress that regardless of which of the two microscopic
mechanisms is the dominant one, the resonant scattering will
reflect the symmetry of the orbital ordering through the
pertubation of the local electronic states at the Mn$^{3+}$ sites.
We believe this is true even though the d orbitals are not
directly involved as intermediate states in the resonant process.
In particular, in terms of the Jahn-Teller distortion considered
above, the orientation of the $e_g$ orbitals and the oxygen motion
reflect the same order parameter. It follows that the peak
positions and widths determined in the x-ray experiments measure
the orbital periodicity and correlation lengths, respectively.
However, it still remains to interpret the x-ray peak intensities
on an absolute scale, which will require additional calculations.
>From this perspective, we think of the resonant scattering as
Templeton scattering arising from the anisotropic charge
distribution induced by orbital ordering. Its basic properties,
for example the polarization and azimuthal dependence, are then
determined by the anisotropy of the susceptibility tensor--which
in the dipole approximation is a second rank tensor.

Working in a linear polarization basis, with  $\sigma$
polarization perpendicular to the scattering plane and $\pi$
parallel, it is easy to show that for a  $\sigma$ incident beam,
the resonant cross-section for an orbital reflection of the type
considered here, does not give rise to a $\sigma^{'}$ polarized
scattered beam. That is, for any azimuthal angle $\psi$,
I$_{\sigma \rightarrow \sigma^{'}}$=0. In the rotated $\pi^{'}$
channel, it may be further shown that:

\begin{equation}
I^{res}_{\sigma \rightarrow \pi^{'}}(\psi) =
\frac{A^4\Delta^2\sin^2\psi}{[\Gamma^2 + 4(2\Delta-x)^2][\Gamma^2 + 4(\Delta + x)^2]}
\end{equation}
where x=($\omega - \omega_o$). This simple model thus predicts
that the scattering is all of the $\sigma \rightarrow \pi$ type
and that it has a twofold azimuthal symmetry, with zeros coming
when the incident polarization is parallel to the c-axis. The
azimuthal angle characterizes rotations of the sample about the
scattering wavevector, and is defined to be zero when the c-axis
is perpendicular to the scattering plane (figure 4). Although a
detailed analysis of the data requires a more sophisticated
treatment, this model captures many of the essential elements of
the experimental results, as will be shown below.

It's worth adding that there should also be nonresonant scattering
from an orbitally ordered material. However, for the structures
shown in Figure 3, and the (0,k,0) reflections, the charge density
is arranged symmetrically, and the scattering is zero.  It is
nevertheless still possible that nonresonant charge scattering can
arise at the orbital wavevector from lattice modulations
accompanying the orbital ordering, and such a modulation has been
observed in Pr$_{1-x}$Ca$_x$MnO$_3$ (see e.g.,
\cite{Jirak85,Yoshizawa95,Kiryukhin97}) and in
La$_{0.5}$Ca$_{0.5}$MnO$_3$\cite{Radaelli97,Radaelli99,Wang_sub,Chen90}.
In fact, one result of the present work is the observation of a
longitudinal component of this modulation for the x=0.4 and 0.5
samples, which we will also discuss below.

\section{Results}
\subsection{Diffraction Pattern}

Scans of the resonant scattering intensity versus momentum
transfer along the (0,k,0) direction are shown for the x=0.25 and
x=0.4 samples, respectively, in Figures 5a and 5b.  Results for
the x=0.5 sample are similar to those shown for x=0.4, and are not
shown. In each case, the samples were cooled below their ordering
temperatures, to T=300 K and T=30 K for the x=0.25 and the x=0.4
samples, respectively. The intensities are plotted versus k in
counts per second and shown on a logarithmic scale. Twinning
within the ab-plane was observed in all three samples. It is
visible as a peak splitting in low momentum-transfer resolution
scans, such as shown in Figure 5a for the x=0.25 sample. This
splitting is not observed in high-resolution scans, for which the
resolution volume never overlaps the second peak (see Figure 5b).

The large peaks falling at k=2,4 in both scans in Figure 5
correspond to bulk allowed Bragg reflections expressed in
orthorhombic units. Their intensities were obtained using Al
absorbers, and should be considered qualitative. Referring to the
x=0.25 sample (Figure 5a), the peaks at k=1,3 correspond to
orbital ordering with the periodicity defined in figure 3a.  Count
rates of ~400/sec were obtained at the (0,1,0) reflection on the
NSLS X21 Wiggler beamline. For the x=0.4 sample, the peaks at
k=0.5, 1.5 and 2.5 correspond to orbital ordering, while those at
k=1,3 correspond to charge ordering, both with the periodicities
defined in Figure 3b. Typical count rates for this sample obtained
at the NSLS bending magnet beamline X22C reached 1500/sec at the
(010) reflection and 3000/sec at the (0,1.5,0) reflection.
Considering the many differences between the two beamlines and the
geometries employed, we have not attempted to make quantitative
comparisons of the intensities. The origins of the peaks at
k$\approx$0.65 and 1.4 are unknown. Both peaks persisted in the
diffraction pattern above the charge and orbital ordering
temperatures, however, and were not studied further.

It's clear from the figure that the wavevectors for charge and orbital
ordering in all three samples are simply commensurate with the lattice,
and independent of concentration. Further, the measured peak positions
are all consistent with the periodicities proposed in Figure 3 for the
different orbital and charge ordered structures.  The temperature
dependence of the charge and orbital order wavevectors will be discussed
further in Section Vc. below.

\subsection{  Resonant Scattering}
\subsubsection{Orbital Wavevector}

Figure 6 shows the energy dependence of the scattering at the
(100) orbital wavevector of the x=0.25 sample as the incident
x-ray energy is tuned through the Mn K absorption edge (6.539
keV). These data were obtained with a Si(111) analyzer on the
CMC-CAT undulator beamline 9ID at the APS. A large resonant signal
is visible at $\hbar\omega$=6.547 keV, reaching more than 20,000
counts per second near the edge.  In addition, there are two
smaller peaks at higher energies ($\hbar\omega$=6.56 and 6.575
keV), and one below (at $\hbar\omega$=6.534 keV). The inset shows
the lower energy peak in more detail. No signal was observed at
energies 100 eV above or below the absorption edge, implying that
only pure resonant scattering was present in this sample.
Polarization analysis (performed on bending magnet beamline X22C
at the NSLS) suggests that the scattered signal is predominantly
$\pi$ polarized, consistent with a rotation of the incident linear
polarization from being perpendicular to the diffraction plane,
$\sigma$, to lying within the diffraction plane, $\pi$. It should
be added that all of the data shown here, and in Figures 7, 10 and
11 below, were obtained at an azimuthal angle $\psi=90^{\circ}$.
As a function of azimuthal angle, the resonant intensity observed
for the x=0.25 sample takes maxima at $\psi$=90 and 270$^{\circ}$,
and minima at 0 and 180$^{\circ}$. These results are all similar
to those obtained previously at the orbital wavevector of
LaMnO$_3$, including the 4-peaked fine structure in the energy
dependence, an identical polarization and azimuthal dependence,
and the absence of nonresonant scattering away from the edge
\cite{Murakami214,Murakami_unpub}.

Figures 7a and 7b show the energy dependence of the scattering at
the (0,1.5,0) and (0,2.5,0) orbital wavevectors of the x=0.4
sample, again as the incident x-ray energy is tuned through the Mn
K absorption edge. These data were obtained on the X22C bending
magnet beamline at the NSLS, and explicitly resolve the
polarization. The closed circles show the $\sigma \rightarrow \pi$
scattering, and the open circles show the $\sigma \rightarrow
\sigma$ scattering. Although the fine structure in figure 7a is not as clearly
resolved as in the x=0.25 sample, the main features of the $\sigma
\rightarrow \pi$ scattering are similar, including a pronounced
resonance peak at 6.547 keV and a weaker peak at 6.57 keV. In
contrast to the $\sigma \rightarrow \pi$ scattering, the $\sigma
\rightarrow \sigma$ scattering shows a double peaked structure
with a pronounced dip at the absorption edge. This is strongly
reminiscent of the behavior of charge scattering, and suggests the
presence of a lattice modulation with the orbital wavevector.  The
fact that the $\sigma \rightarrow \sigma$ intensity does not fall
off at lower x-ray energy, but instead continues above background,
is further evidence of a significant nonresonant signal as would
be produced by such a modulation.  Lattice modulations associated
with the CE type structure have been observed before in
Pr$_{1-x}$Ca$_x$MnO$_3$ e.g.\cite{Jirak85,Jirak2,Yoshizawa95,Kiryukhin97}
and also in
La$_{0.5}$Ca$_{0.5}$MnO$_3$\cite{Radaelli97,Hirota2000,Hatton99}.
In the latter compound, the structure was solved, and the
modulation found to be purely transverse.  Such a modulation,
however, is inconsistent with the present results.  We will return
to this point below.

A broader ranged energy scan of the (0,2.5,0) orbital wavevector
is shown for the x=0.4 sample in Figure 8. These data were taken
with a Ge(111) analyzer, and are plotted versus energy for
several different azimuthal angles. The use of a Ge analyzer
implies that both $\sigma \rightarrow \sigma$ and $\sigma
\rightarrow \pi$ components are detected, and that their
intensities add. Referring to the scan for $\psi=85^{\circ}$, the
basic features shown in Figure 7b are reproduced, although the
$\sigma \rightarrow \sigma$ scattering clearly dominates the
signal. Below the Mn absorption edge, the observed scattering is
approximately constant until it reaches the Pr L$_2$ absorption
edge energy at 6.43 keV. There the intensity again shows a dip,
primarily as a result of the increase in the absorption.
Qualitatively similar results were obtained for the x=0.5 sample.

It's clear from Figure 8 that except for a variation of the
overall intensity, no new features are introduced as a function of
azimuthal angle. A quantitative study of the azimuthal dependence
of the orbital scattering at (0,2.5,0) is shown in Figure 9, in
which the maximum resonant intensity in the $\sigma \rightarrow
\pi$ geometry was recorded versus azimuthal angle for rotations
over 180$^{\circ}$. The data have been normalized by the average
intensity of the (0,2,0) and (0,4,0) fundamental Bragg reflections
to correct for small variations due to sample shape. Again,
$\psi$=0 corresponds to a configuration in which the c-axis is
perpendicular to the diffraction plane. In contrast to a normal
charge reflection, for which the intensity is independent of the
azimuthal angle, the resonant scattering exhibits a characteristic
oscillation with a two-fold symmetry. The intensity approaches
zero when $\psi$=0 and 180$^{\circ}$, consistent with the $\sigma
\rightarrow \pi$ polarized component of the resonant scattering in
the x=0.25 sample. The solid line in Figure 8 is a fit to the form
$A\sin^2\psi$, as predicted in equation 2.

To summarize: In all three samples, we find a pure resonant signal
in the $\sigma \rightarrow \pi$ channel at the appropriate orbital
wavevector with the dominant peak located near the Mn K absorption
edge. Additional fine structure is observed both above and below
the absorption edge. The $\pi$-resonant scattering has the
characteristic azimuthal dependence, varying as $\sin^2\psi$,
where $\psi$ is the azimuthal angle. These results are identical
to what has been observed previously in
LaMnO$_3$\cite{Murakami113} and
La$_{0.5}$Sr$_{1.5}$MnO$_4$\cite{Murakami214}. In the x=0.4 and
0.5 samples, there is in addition a $\sigma \rightarrow \sigma$
component of the scattering at the orbital wavevector with both
resonant and nonresonant parts. The non-resonant component lacks
any azimuthal dependence and is consistent with normal charge (or Thomson)
scattering. The x=0.25 sample lacks a $\sigma \rightarrow \sigma$
component to within the detection limits of the experiment, as was
also the case in LaMnO$_3$.

We associate the dominant, resonant peak, which occurs in the
$\sigma \rightarrow \pi$ channel of all three samples with the
electric dipole transition coupling 1s and 4p states, as discussed
in section IV. Recall that the sensitivity to orbital ordering may
be thought of, qualitatively, as arising from a splitting of the
Mn 3d states, either through the Jahn-Teller distortion of the
oxygen atoms or through a Coulomb interaction. In either case, the
existence of a dipole resonance in the $\sigma \rightarrow \pi$
channel, and the observed azimuthal dependence, are consistent
with theoretical predictions.  We, therefore, interpret the
observed resonant scattering as Templeton scattering induced by
the orbital ordering, just as previously concluded for LaMnO$_3$
and La$_{0.5}$Sr$_{1.5}$MnO$_4$\cite{Murakami214,Murakami113}. To
explain the additional fine structure both above and below the
main peak, however, requires a more sophisticated treatment.
Elfimov et al. \cite{Elfimov99}, in particular, have performed
band structure calculations for LaMnO$_3$ with a LSDA+U-type
approach, and shown that the fine structure above the absorption
edge reflects the 4p density of states after hybridization of the
central Mn 4p and surrounding O 2p states. They show further that
the higher energy peaks originate predominantly from the
Jahn-Teller distortion of the oxygens, and not from direct Coulomb
interactions. In contrast, the small peak about 13 eV below the
white line is associated with the intersite 4p-3d hybridization
of the central and neighboring Mn ions via the
intervening 0 2p states.  Although quantitative comparisons remain
to be made, the qualitative agreement between these predictions
and the observed spectra is good, and offers a natural description
of the experimental results. In this regard, it should be added
that Ishihara {\em et al.} \cite{Ishihara_condmat} have also
carried out cluster calculations of the resonant cross-section in
LaSr$_{2}$Mn$_2$O$_7$, assuming an intra-site 3d-4p Coulomb origin
of the 4p splitting. By including band effects, they also were
able to produce qualitatively similar fine structure above the Mn
K edge. Thus, we are not able to distinguish a possible Coulomb
origin of the resonant peak from a Jahn-Teller origin on the basis
of our experiments ---however, the additional fine structure at
higher photon energies appears to result from band effects in both
approaches.

It's worth noting that the energy of the pre-edge feature (see
inset, fig. 6) corresponds to that of the Mn 3d states.   Several
groups have shown that this feature is highly sensitive to the 3d
orbital
occupancy\cite{Elfimov99,Benfatto99,Fabrizio98,Takahashi2000}, but
not to the Jahn-Teller distortion (oxygen motion)-in contrast with
the main-edge feature.  In fact, states of both d- and p-like
symmetry (with respect to the central ion) exist at the pre-edge
energy, with significantly more weight in the d-like states
\cite{Takahashi2000}, suggesting that both dipole and quadrupole
processes should contribute. Takahashi, {\em et al.} estimate that
the total intensity, enhanced by interference with the main-edge
processes, is about 1\% of the main edge. In our experiment, we
find that  the pre-edge intensity is about 5\% of the main edge
intensity. In addition, we observed only a single feature, in
contrast with the two features, separated by 3 eV, predicted by
Takahashi, {\em et al.} \cite{Takahashi2000}.  Our energy
resolution for these data was ~ 1.5 eV. In principle, the pre-edge
feature could exhibit different azimuthal \cite{Takahashi2000} and
temperature dependences from those of the main-edge feature. In
our studies of the azimuthal and temperature dependence however,
we found no difference between the pre-edge and main-edge
behaviors, to within errors.

We note, in passing, that in V$_2$O$_3$ the resonant ion is not in
a center of inversion symmetry and that, therefore, dipole
transitions are allowed directly into the d-band of that material.
This gives rise to a large pre-edge feature which has also been
used to study orbital order \cite{Fabrizio98,Paolosini99} by x-ray
scattering techniques.

We now turn to the lattice modulation observed at the orbital
wave-vector in the $\sigma \rightarrow \sigma$ channel. Such
modulations have been observed before in CE-type structures, and
in particular in Pr$_{1-x}$Ca$_x$MnO$_3$ by
neutron\cite{Jirak85,Jirak2,Kiryukhin97} and non-resonant x-ray
scattering\cite{Kiryukhin97,Cox98}. Similar results have also been
obtained in La$_{0.5}$Ca$_{0.5}$MnO$_3$\cite{Radaelli97} at
($h,\frac{k}{2}-\epsilon ,2n$), where $h\neq 0$ and $\epsilon$,
the incommensurability, is weak (we have converted to Pbnm
settings to be consistent with the rest of the present paper). The
latter structure was solved by x-ray powder diffraction and a
purely transverse modulation of the Mn$^{4+}$ sites was deduced,
wherein the Mn$^{4+}$ sites are displaced along the a-direction
with a periodicity equal to the orbital periodicity. All of the
orbital superlattice peaks observed to date in the
Pr$_{1-x}$Ca$_x$MnO$_3$ system have also had a significant a-axis
component and a similar distortion has been assumed
\cite{Jirak85,Jirak2}.

In the present case, the $\sigma\rightarrow\sigma$ scattering
observed at $(0,k+\frac{1}{2},0)$ requires a longitudinal b-axis
component of the modulation.  This follows from the
small-displacement limit of the x-ray intensity, which varies as
$Q\cdot\delta$ to leading order for displacements of the form
$\delta\sin(\tau.R)$.  For x=0.4 and 0.5, we did not examine
reflections of the form {(h/2,0,l)} with {l} $\neq$ 0 and cannot
draw conclusions about displacements in other directions, however,
we have recently performed limited studies on an x=0.3 sample in
which the orbital reflections around (0,2,0) and (2,2,0) were
studied. The non-resonant scattering was observed to be
significantly stronger in the vicinity of the (220) consistent
with a larger transverse displacement and a small longitudinal
modulation\cite{Nelson}.  We will return again to the lattice
modulation shortly.

It is also worth commenting on the differences between the
La$_{0.5}$Ca$_{0.5}$MnO$_3$ structure and the present case. In
La$_{0.5}$Ca$_{0.5}$MnO$_3$, a temperature dependent
incommensurability was observed at the orbital wavevector
\cite{Elfimov99,Chen99}, whereas in the present case the
scattering is strictly commensurate. The source of the
incommensurability in the former material is believed to be an
ordered array of domain walls - discommensurations - separating
regions of commensurate order.  As is discussed below, we observe
domain walls at a similar spacing in Pr$_{0.5}$Ca$_{0.5}$MnO$_3$.
Thus, the main difference between the two structures appears to be
that the domain walls are ordered in La$_{0.5}$Ca$_{0.5}$MnO$_3$,
but disordered in Pr$_{1-x}$Ca$_x$MnO$_3$. It is an interesting
question as to why this is so.

\subsubsection{Charge Order Wavevector}

Figures 10a and 10b show the energy dependence of the scattering
at the (0,1,0) and (0,3,0) charge order wavevectors of the x=0.4
sample, as the incident x-ray energy is tuned through the Mn K
absorption edge. These data were obtained at the NSLS beamline
X22C and are polarization resolved. The open circles show the
$\sigma \rightarrow \sigma$ scattering and the closed circles the
$\sigma \rightarrow \pi$ scattering. No signal was obtained in the
$\sigma \rightarrow \pi$ channel, to within the detection limits
of the experiment. In the $\sigma \rightarrow \sigma$ channel the
(0,1,0) reflection has a shoulder at lower energy which rises to a
resonant peak at 6.544 keV. This is followed by a dip and another
smaller peak centered near 6.58 keV. The profile of the (0,3,0)
reflection shows a resonant peak at slightly higher energy (6.546
keV) relative to the (0,1,0) reflection, and the additional
structure appears inverted.  This is a clear signature of an
interference process, involving the resonant and nonresonant
contributions to the charge order scattering. The nonresonant
scattering may, in principle, result from the valence modulation
itself (which is weak), or from an accompanying lattice
modulation, or both. The resonant scattering arises from the
anomalous parts of the Mn$^{3+}$ and Mn$^{4+}$ scattering factors,
which are distinct.

For comparison, the energy dependence of the scattering at the
charge order wavevector (010) of the x=0.5 sample is shown in
Figure 11. The data were obtained without an analyzer, and so
include both the $\sigma \rightarrow \pi$ component and any
$\sigma \rightarrow \pi$ component.  The lineshape is nearly
identical to that obtained for the x=0.4 sample at the same
reflections.

A series of broader ranged energy scans of the (0,3,0) reflection
taken at various azimuthal angles for the x=0.4 sample is shown in
Figure 8. The basic features noted in Figure 10 are reproduced
there. In addition, there is a dip in the scattering at 6.44 keV,
which reflects the Pr L$_2$ absorption edge. We believe that the
peak at 6.64 keV in the scan at $\psi$=0 arises from multiple
scattering and can be ignored. Except for a decrease of the
resonant intensities, very little else changes in these spectra as
a function of azimuthal angle, similar to what was observed for
orbital ordering.

A quantitative study of the dependence of the charge order
scattering on azimuth at the (0,3,0) reflection is shown in figure
9. The filled squares record the behavior of the maximum resonant
intensity obtained at 6.546 keV, while the open squares give the
intensity off resonance at 6.47 keV. As before, these data have
been normalized by the average of the (020) and (040) fundamental
Bragg intensities. In contrast to the nonresonant charge order
scattering, which is flat as expected, the resonant charge order
scattering exhibits a pronounced azimuthal dependence, with the
same sin-squared behavior as observed above for the resonant
orbital scattering.

The explanation of this azimuthal dependence follows simply from
the structure factor of (0,k,0) charge-type peaks:
\begin{equation}
f(0,2n+1,0) = f_{3A}^{n=1} + f_{3B}^{n=-1} - 2f_{4+},
\end{equation}
where f$_{3A(B)}^{n=1(n=-1)}$ are the atomic form factors for the Mn$^{3+}$
ions on the A(B) orbital sub-lattice, and n=1,-1 refers to figure 4.  
As emphasized above, these
quantities are second rank tensors near  resonance.  We take
f$_{4+}$ to be spherically symmetric and we write
\begin{equation}
f_{3A} =\left( \begin{array}{ccc}
                       a&0&0 \\
                       0&b&0 \\
                       0&0&b
                \end{array}
                \right) ,
~{\rm and}~f_{3B} =\left( \begin{array}{ccc}
                        b&0&0 \\
                        0&a&0 \\
                        0&0&b
                \end{array}
                \right) .
\end{equation}

where f$_{3A}$ and f$_{3B}$ have the symmetries $3x^2-r^2$ and
$3y^2-r^2$, respectively. Calculating the polarization and
azimuthal dependence of the scattering,
\begin{equation}
I = \left| \vec{\epsilon}_i \cdot \tilde{F}_{0k0}(\psi)\cdot
\vec{\epsilon}_f \right | ^2,
\label{azi_chargeorder}
\end{equation}
we find a two fold azimuthal dependence in the $\sigma \rightarrow
\sigma$ channel, with zeroes at $\psi=0$ and 180$^{\circ}$, as
observed. In addition, equation \ref{azi_chargeorder} predicts a
$\sigma \rightarrow \pi$ component with a four-fold symmetry (with
zeros at 0,90,180,270$^{\circ}$) although with a significantly
smaller intensity (which we were unable to observe). We remark
that it is the anisotropy of the structure factor f(0,2n+1,0) that gives rise to the
observed azimuthal dependence. In this sense, the ``charge-order''
reflection has some orbital character.

In order to model the energy dependence of the charge order
scattering, we write the structure factor as a sum of the
scattering factors of the Mn$^{3+}$ and Mn$^{4+}$ ions within the
unit cell, including both resonant and nonresonant terms for each.
Here, for simplicity, we neglect the tensor character discussed
above, and treat the form factors as (complex) scalars.  Figure
12a shows schematic forms of the real and imaginary parts of the
Mn$^{3+}$ scattering factor, plotted versus incident photon
energy. As our intention in the following was to gain a
qualitative understanding of the energy lineshape, the parameters
have been chosen for convenience of illustration, and do not
represent the actual properties of Mn. The Mn$^{4+}$ scattering
factor is then obtained from that of Mn$^{3+}$ by shifting the
absorption curve by 2eV, following Murakami, {\em et al.}
\cite{Murakami214}.  Assuming that the Mn displacements  $\delta$
are along the b-axis with modulation wavevector (0,k,0) in the
x=0.4 and 0.5 samples (see inset Figure 12b), the structure factor
may be written:

\begin{equation}
f(0,k=odd,0) = f_{3+}e^{i\pi k\delta} - f_{4+}e^{-i\pi k\delta},
\end{equation}

where  $\delta$ is a displacement parameter (in units of b, the lattice
constant). This reduces to the simple form ($f_{3+}-f_{4+}$) when
$\delta$=0, as expected.

A plot of the energy dependence of the intensity predicted by this
model for the (0,1,0) and (0,3,0) reflections is shown in Figure
12b for $\delta$=0 and including nonresonant scattering arising from 
oxygen motion. It reproduces the basic features of the lineshapes in Figure
10, including the interference.  The lineshapes also resemble
those generated by Murakami et al.\cite{Murakami214} for
La$_{0.5}$Sr$_{1.5}$MnO$_4$, using a more quantitative model. 
If the manganese atoms are then allowed to move ($\delta \neq$ 0),
the small shift of the peak maximum of the (0,1,0)
reflection relative to the (0,3,0) reflection is also
described. This is illustrated in Figure 12c.
It is worth noting that a simple transverse
displacement cannot explain the x-ray results for the charge
ordering, just as was noted earlier for the modulation
accompanying the orbital ordering. On the basis of these results,
we conclude that the x=0.4 and 0.5 samples exhibit lattice
modulations with displacements along the b-direction, and produce
scattering at both the charge and orbital wavevectors.  A fuller
understanding of the lattice distortions observed in these
materials will require further experiments.

\subsection{ Temperature Dependence}
\subsubsection{Intensities}

The temperature dependence of the orbital ordering intensity between 10
and $\approx$ 850 K is shown for the x=0.25 sample in Fig. 13.  Each
point represents the peak intensity obtained from a k-scan of the
(0,1,0) reflection, such as is shown in Fig. 5a.  Circles and squares
represent the results of two different runs, one  taken at high
temperatures with the sample heated in an oven and the other at low
temperatures with the sample cooled in a cryostat.  The two data sets
were then scaled to be equal at 300 K.  Referring to Fig. 13, the
orbital intensity is approximately constant (to within 3$\sigma$)
between 10 and 200 K, and decreases gradually with a long tail until
about 850 K -  a surprising result.  High momentum-transfer resolution
scans showed further that the orbital peak widths were independent of
temperature and resolution-limited throughout the ordered phase.  Thus,
the intensities shown in Fig. 13 may be regarded as integrated
intensities.  The resolution-limited behavior also implies that the
corresponding orbital correlation lengths were at least 2000 $\AA$ along
the b-direction.

The temperature dependences of the integrated charge and orbital
ordering intensities of the x=0.4 sample are plotted between 10 and 300
K in Fig. 14a,b. The charge order intensities were obtained at the
(0,3,0) reflection both at resonance using a Ge(111) analyzer and
off-resonance at 6.6 keV.  The latter employed a Cu(220) analyzer. The
orbital intensities were obtained at resonance at the (0,1.5,0)
reflection, where the resonant $\sigma \rightarrow \pi$ scattering
dominates the signal and off resonance at the (0,2.5,0) reflection, with
an incident photon energy of 6.47 keV. Both data sets for the orbital
ordering were obtained using a Ge(111) analyzer.

Referring to the figure, the orbital and charge intensities are
approximately constant between 10 and 130 K, but begin to decrease above
$\approx$130 K.  They drop abruptly to zero at 245 K, consistent with a
first order, or nearly-first-order, transition.  It is clear from the
figure that the temperature dependences of the resonant and nonresonant
intensities are identical for the charge and orbital ordering,
respectively, at least to within the present counting statistics.  This
suggests that the lattice modulation accompanying the charge and orbital
ordering reflects the same order parameter.  The temperature dependences
of the charge and orbital ordering in the x=0.4 sample are also very
similar, as shown directly in an earlier publication
\cite{Zimmermann99}. There we speculated that the charge and orbital
ordering might be linearly coupled.  More recently, Zhang and Wang have
argued on the basis of a Landau theory that the charge and orbital
ordering in Pr$_{1-x}$Ca$_x$MnO$_3$ are quadratically coupled
\cite{Zhang99}, but that the coupling is sufficiently strong that they
have very similar temperature dependences over the range considered
here.

The temperature dependences of the orbital and charge order intensities
of the x=0.5 sample between 10 and 300 K are shown in Fig. 15a,b.  The
orbital ordering intensities were again obtained from k-scans of the
(0,1.5,0) reflection using Ge(111) and Cu(220) analyzer configurations,
with the latter in a $\sigma \rightarrow \pi$ geometry.  The signal
rates in these experiments were weak,($ < $10/sec at T=220K), making a
definitive characterization of the temperature dependence difficult.
Charge order intensities were obtained at the (0,1,0) reflection using a
Ge(111) analyzer.  All of these data were obtained at the Mn K-edge
resonance.

Referring to figures 15a and 15b, the orbital ordering intensity is
approximately constant between 10 and 150 K, and begins to fall
off between 150 and 200 K,  reaching zero at about 250 K.  The
charge order intensity is also constant between 10 and 150 K, but
exhibits a much clearer decrease near the Neel temperature T$_N$ =
170 K.  A similar decrease at T$_N$ was observed for the orbital
intensity of LaMnO$_3$ \cite{Murakami113}, where a correlation was
made between the long-range antiferromagnetic order and the
orbital order.  Although this trend is suggested in each of the
samples studied here, with T$_N$ = 140, 170 and 180 in the x =
0.25, 0.4 and 0.5 samples, respectively, only the data for the
charge ordering of the x = 0.5 sample is convincing.  Despite
repeated attempts to quantify this behavior in the other samples,
we still cannot claim to have measured a definitive link between
the magnetic ordering and the charge and orbital ordering in each
case.

\subsubsection{Correlation Length}

High momentum transfer resolution longitudinal scans of the Bragg,
charge and orbital ordering lineshapes of the x = 0.4 and 0.5
samples are superimposed on each other for comparison in Figure
16a, b.  The data were obtained at low temperatures (10K) in the
ordered phase using a Ge(111) analyzer. Solid lines indicate the
results of scans through the (0,2,0) Bragg peaks; open circles
indicate scans through (0,2.5,0) orbital peaks; and filled circles
give the results obtained for the (0,3,0) and (0,1,0) reflections
of the charge-ordered peaks of the x = 0.4 and 0.5 samples,
respectively. It is clear from the figure that the Bragg and
charge order peaks have similar widths, approximately
corresponding to the momentum-transfer resolution at each Q.  This
implies that the correlation lengths of the structure and of the
charge order are at least 2000 $\AA$ in each case.  The small
differences in width probably reflect the Q-dependence of the
resolution function.  In contrast, the orbital ordering peaks in
both samples are significantly broader then the resolution and
imply smaller orbital domain sizes.

Estimating the correlation length, $\xi$, with $\xi = b/2\pi\Delta
k$, where b is the lattice constant, and $\Delta k$ the half width
of the orbital peaks, we find orbital correlation lengths  $\xi=
320 \pm 10 \AA $ and $160 \pm 10 \AA$ for the x = 0.4 and 0.5
samples, respectively.  We note that in the analysis the charge
order peaks were fitted with Lorentzian lineshapes, while the
orbital ordering peaks were fitted with squared Lorentzians.  When
necessary for deconvolution, the Lorentzian squared resolution
function was used, as derived from the structural Bragg peaks.
These lineshapes were chosen simply for the quality of the fit,
and have not been justified theoretically.

It follows that the x = 0.4 and 0.5 samples do not exhibit long
range orbital order, but instead form a domain state with randomly
distributed antiphase domain boundaries.  A schematic view of such
a domain wall is shown in Fig. 3c. In contrast, the charge
ordering is much more highly correlated.

The presence of an orbital domain state sheds light on recent
neutron diffraction studies of Pr$_{0.5}$Ca$_{0.5}$MnO$_3$
\cite{Kajimoto98,Jirak2}, and powdered
La$_{0.5}$Ca$_{0.5}$MnO$_3$\cite{Radaelli99}. In the former it was
shown that the magnetic correlation length was finite. In
La$_{0.5}$Ca$_{0.5}$MnO$_3$, which also exhibits the CE magnetic
structure with orbital and charge order\cite{Radaelli97}, separate
magnetic correlation lengths were extracted for the Mn$^{3+}$ and
Mn$^{4+}$ magnetic sublattices, with the remarkable result that
they were quite different: $\xi_{3+}^{mag} = 250-450  \AA$ and
$\xi_{4+}^{mag} \geq 2000 \AA$, respectively.  Those authors
proposed antiphase domain walls composed of ``mis-oriented'' $e_g$
orbitals to explain the magnetic disorder of the Mn$^{3+}$
sublattice.  Such domain walls break the magnetic coherence on the
3+ sublattice only, as long as the charge order is preserved (fig.
3c).  Our results strongly suggest that this interpretation is
correct, and that we have observed the antiphase domains directly
in the orbital reflection (although in a different material).
These results taken together suggest that such orbital domain
states are common to these systems - at least in this range of
doping.  Note that these domains are believed to be static, and do
not correspond  to the (dynamic) orbital fluctuations inferred
from magnetic neutron diffraction investigations of the
ferromagnetic spin fluctuations in Pr$_{1-x}$Ca$_x$MnO$_3$, which
disappear below T$_N$ \cite{Kajimoto98}.

The temperature dependences of the orbital widths of all three
samples are compared between 10 and 300 K in Fig. 17.  At 10 K,
the (undeconvolved) widths of the x = 0.25, 0.4 and 0.5 samples
are about 0.0003, 0.0016 and 0.003 r.l.u., respectively, giving
the correlation lengths noted above.  There is a clear decrease of
orbital domain size as the doping increases from x = 0.25 to 0.5.
A possible explanation for difference between the x = 0.4 and 0.5
samples follows from the fact that the x = 0.5 sample is closer to
tetragonal than the x = 0.4 sample. Specifically, $\zeta$(x = 0.5)
= 2(a-b)/a+b = 1.48 $\times 10^{-3}$ compared $\zeta$(x = 0.4) =
4.23 $\times 10^{-3}$ at room temperature \cite{Jirak85}. In the
more tetragonal sample, the a and b domains are more nearly
degenerate and the energetic cost of a domain wall is, therefore,
reduced \cite{Millis_private}. With regard to the difference
between the x = 0.25 sample and the x = 0.4 and 0.5 samples, we
emphasize that the increased doping introduces charge ordering
into the orbital lattice and changes the orbital periodicity
(compare Figures 3a, b), and it is not clear that the two
situations may be directly compared.

\subsubsection{Lattice Constants and Wavevectors}

The temperature dependence of the b-axis lattice constants between
10 and 300 K is shown for the x = 0.4 and 0.5 samples in Fig. 18.
These data were obtained from measurements of the (0,2,0) and
(0,4,0) bulk Bragg peaks.  As shown in the figure, there is an
abrupt change in lattice parameter of both samples at the onset of
charge and orbital ordering near 250 K.  This is consistent with
the formation of both a cooperative Jahn-Teller distortion and a
longitudinal lattice distortion as discussed in section Vb above.
In the x = 0.4 sample, the lattice constant is approximately
constant below T$_o$ = 245 K, and then decreases slightly at the
magnetic ordering temperature (T$_N$ = 170 K). The lattice
constant of the x = 0.5 sample increases slightly below T$_o$, and
then levels off below T$_N$.  These changes at T$_o$ and T$_N$ are
greatest for the x = 0.4 sample, which also exhibits the longer
ranged orbital order.  This implies that both the orbital and
magneto-striction may be partially compensated at domain
boundaries, which occur more frequently in the x = 0.5 sample.
The data obtained for the b-axis lattice constant of the x=0.5 sample
are in quantitative  agreement with results of Jirak et al. \cite{Jirak2}
on a powdered sample.

The temperature dependences of the charge and orbital wavevectors
are plotted for the x = 0.25, 0.4 and 0.5 samples in Fig. 19.  All
wavevectors remain commensurate with the lattice and locked to
either (0,1,0) or (0,1/2,0), throughout the ordered phase,
independent of temperature to within 0.002 r.l.u. .  This is in striking disagreement with
the results of electron diffraction studies by Chen, et al.
\cite{Chen99} and Jirak et al. \cite{Jirak2},  in which the orbital order wavevectors were found
to change by as much as 30\%. Indeed, Jirak et al. suggest the 
possibility of a Devil's staircase in the temperature dependence 
of the charge order wavevector. The results of Chen at al. \cite{Chen99}
are schematically
shown by the crosses in Fig. 19. The origin of the difference
between these results and ours is not understood.

\subsubsection{Scattering Above the Charge and Orbital Ordering Temperature}

The behavior of the charge and orbital ordering in the vicinity of
the phase transition at T$_o$ = 245 K is illustrated for the x =
0.4 sample in Fig. 20. Longitudinal scans were taken upon warming
of the (0,3,0) reflection in a $\sigma \rightarrow \sigma$
geometry and of the (0,2.5,0) reflection in a $\sigma \rightarrow
\pi$ geometry.  The peak intensities are plotted on a logarithmic
scale between 200 and 280 K in Fig. 20a. Measurable charge order
fluctuations are observed at much higher temperatures above 245 K
than for orbital fluctuations, with weak charge-order  scattering
still present at 280 K. In contrast, the orbital fluctuations fall
off more quickly, and have disappeared by 260 K, at least to
within the present counting statistics.  The corresponding peak
widths are considerably narrower for the charge order than for the
orbital order (Fig. 20b), that is, the correlation lengths of the
charge order fluctuations are significantly longer than those of
the orbital order fluctuations at any given temperature above
T$_o$ = 245 K.  In this regard, it is worth noting that the
correlation length of the charge order must be at least as long as
that of the orbital order since the orbital unit cell is defined
on the charge order lattice.

The picture these data present for the charge and orbital ordering
transition in the x = 0.4 sample is one in which the transition
proceeds via local charge order fluctuations which grow as the
transition is approached from above.  Long-range charge order is
nucleated at the transition temperature. Orbital fluctuations are
induced by the charge order fluctuations and become observable
only close to the transition. The coupling mechanism has yet to be
fully elucidated, however, a quadratic coupling has been suggested
by Zhang and Wang\cite{Zhang99}, on the basis of a Landau
theory as noted above.

The temperature dependences of the charge and orbital order half
widths are plotted between 180 and 300 K for the x = 0.5 sample in
Fig. 21a,b. The orbital scattering was characterized in a low
resolution mode with a Cu(220) analyzer in the $\sigma \rightarrow
\pi$ geometry.  The widths show a clear increase just below the
transition, however, the signal is extremely weak and disappears
about 1K above it.  The charge order scattering was characterized
using a higher-resolution Ge(111) analyzer. Its width also
increases near the transition, however, is always smaller than the
smallest value exhibited by the orbital scattering in the ordered
phase. Thus, the length scale of the charge order fluctuations
exceeds that of the orbital fluctuations over the narrow
temperature range in which both exist.  In this sense the data for
the x = 0.5 sample are consistent with the results for the x = 0.4
sample, though they are neither as convincing nor as clean.
Definitive conclusions will require further experiments at a more
intense beamline. Preliminary results of such experiments support
the suggestions made here. It is worth adding that the orbital
scattering observed in the x = 0.25 sample was not observed to
broaden, within experimental errors, at any temperature.

Finally, we note that the narrowing of the orbital reflection in
the x=0.5 sample below T$_o$ is consistent with the behavior
previously seen in a number of diverse systems for which disorder
prevents the particular order parameter reaching its long-range
ordered ground state. Examples include magnetic order in dilute
antiferromagnets in applied fields (see for example\cite{Hill93}
and doped spin Pierels systems\cite{Wang99}.  In these two
examples, the common phenomenology appears to be rapidly
increasing time scales as the transition is approached, such that
the system is not able to fully relax and reach equilibrium. As
the temperature is reduced further, equilibration times become
longer than experimental measuring times and the system freezes in
a metastable glass-like state.  It is possible that similar
phenomena underlie the behavior observed here.  As noted above,
the disorder is greater in the x=0.5 sample.  In this regard, we
note that the lineshape of the orbital reflection, a Lorentzian
squared, is the same as that observed in the dilute
antiferromagnets. It may be derived from exponentially decaying
real space correlations.

\section{Magnetic Field Dependence}

An intriguing property of the perovskite manganites is the
existence of colossal magnetic resistance behavior in an applied
magnetic field.  The transition to a metallic phase involves the
delocalization of the Mn e$_g$ electrons, which leads to the
destruction of static charge and orbital ordering.  That this
transition can be driven by a magnetic field in
Pr$_{1-x}$Ca$_x$MnO$_3$ was first demonstrated by Tomioka, et al
\cite{Tomioka96}. It is an interesting question whether the same
phenomenology of the transition discussed above applies when the
transition is driven at fixed temperature and the fluctuations are
activated by a magnetic field. We have carried out studies of the
transition at two temperatures, T = 30 and 200 K, with critical
fields of H$_o$ = 6.9(1) and 10.4 T, respectively
\cite{Tomioka96}. The field dependence of the charge and orbital
intensities of the x = 0.4 sample taken at T = 30 K are
illustrated in Fig. 22.  The intensities of the (0,3,0) and
(0,2.5,0) reflections exhibit identical field dependences below
the transition.  Above the transition, the charge order
fluctuations are markedly stronger than the orbital fluctuations.
A similar behavior was also observed at T = 200 K, i.e., charge
order fluctuations were observed at fields for which orbital
fluctuations were no longer observable (see inset, Fig. 22). From
this it appears that the disorder transition is driven by charge
order fluctuations for both the temperature- and field- driven
cases. We note that as a result of experimental constraints it was
only possible to measure the charge and orbital ordering at a
photon energy of 8 keV in an applied magnetic field.  The
corresponding nonresonant intensities are sufficiently weak above
6.5T that it was not possible to obtain reliable values of the
half widths.

\section {Conclusions}

This paper describes a systematic study of the charge and orbital
ordering in the manganite series Pr$_{1-x}$Ca$_x$MnO$_3$ with x =
0.25, 0.4 and 0.5.  The temperature dependence of the charge and
orbital ordering was characterized in each sample, including the
correlation lengths, wavevectors and lattice constants.  It was
found that the charge and orbital ordering are strictly
commensurate with the lattice in all three samples, in striking
contrast to electron diffraction studies of samples with x = 0.5.
We do not understand the origin of these discrepancies. High
Q-resolution scans revealed that long-range orbital order is never
established in the x = 0.4 and 0.5 samples, a result that makes
connections with the observation of finite Mn$^{3+}$ magnetic
domains in neutron  diffraction studies of
Pr$_{0.5}$Ca$_{0.5}$MnO$_3$ and La$_{0.5}$Ca$_{0.5}$MnO$_3$. We
suggest that the orbital domains observed here represent the
antiphase magnetic domain walls responsible for destroying the
Mn$^{3+}$ magnetic order. The observation of such an orbital
glass-like state in two such materials suggests that such
phenomena may be common in manganites, at least those with the
CE-type charge/orbital structure. Above the charge/orbital
ordering temperature T$_o$, we found that the charge order
fluctuations are more highly correlated than the orbital ordering
fluctuations.  This suggests that charge order drives orbital
order in these systems.  In addition, we reported on the
destruction of charge and orbital ordering in the x = 0.4 sample
by the application of a magnetic field.  A similar phenomenology
was found for increasing field as was found for increasing
temperature. Finally, the polarization and azimuthal dependence of
the resonant charge and orbital ordering at the K edge is found to
be qualitatively consistent with the predictions of a simple model
in which the 4p levels are split in the orbitally ordered phase.
The experiments do not distinguish among more sophisticated
treatments of the electronic structure, whether in atomic or band
limits.  A new result is the discovery of a longitudinal lattice
modulation in the x = 0.4 and 0.5 samples with scattering at both
the charge and orbital wavevectors. The presence of such a
component of the modulation represents a refinement of earlier
models in which the displacement was assured to be purely
transverse, following work on La$_{0.5}$Ca$_{0.5}$MnO$_3$. Similar
results have been reported in Pr$_{0.5}$Ca$_{0.5}$MnO$_3$,
however, there the displacement is transverse.

\section{ Acknowledgements}

We acknowledge helpful conversations with S. Ishihara, D.J. Khomskii, 
S. Maekawa, A.J. Millis, and G. A. Sawatzky. The work at Brookhaven, both in the
Physics Department and at the NSLS, was supported by the U.S. Department
of Energy, Division of Materials Science, under Contract No.
DE-AC02-98CH10886, and at Princeton University by the N.S.F. under Grant
No. DMR-9701991. Support from the Ministry of Education, Science and
Culture, Japan, by the New Energy and Industrial Technology Development
Organization (NEDO), and by the Core Research for Evolution Science and
Technology (CREST) is also acknowledged. Work at the CMC beamlines is
supported, in part, by the Office of Basic Energy Sciences of the U.S.
Department of Energy and by the National Science Foundation, Division of
Materials Research.  Use of the Advanced Photon Source was supported by
the Office of Basic Energy Sciences of the U.S. Department of Energy
under Contract No. W-31-109-Eng-38.

\bibliography{PCMO}
\bibliographystyle{Prsty}

\figure{
Schematic structure of Pr$_{1-x}$Ca$_x$MnO$_3$. Small spheres correspond
to oxygen, and large spheres to Pr or Ca. The Mn atoms are at the center
of the octahedra. Solid lines show the orthorhombic unit cell used in
this paper.}

\figure{ Composition-temperature phase diagram of
Pr$_{1-x}$Ca$_x$MnO$_3$ in zero magnetic field (following
reference \cite{Jirak85}).}

\figure{
Schematic of the charge, orbital and magnetic order in
Pr$_{1-x}$Ca$_x$MnO$_3$. Filled circles represent Mn$^{4+}$ ions, shaded
figure-eights represent Mn$^{3+}$ ions, and the arrows indicate the
in-plane components of the magnetic ordering. Solid lines show orbital
order unit cell; dashed lines show the charge order unit cell. (a)
Proposed orbital ordering for x=0.25, (b,c) Charge and orbital order for x=0.4
and 0.5., and (c) shows an orbital antiphase domain wall.}

\figure{
Upper panel: Schematic view of the orbital in the a-b plane of the
LaMnO$_3$. Lower panel: Schematic energy level diagram of Mn
4p$_{x,y,z}$ in the orbitally ordered state, for the two orbital
sublattices.}

\figure{
Upper: Scan along (0,k,0) of the x=0.25 sample at T=300 K at the 
resonance Energy (E=6.547 keV). Lower: Scan
along (0,k,0) for the x=0.4 sample at T=30 K (E=8 keV).}

\figure{ Intensity plotted versus incident photon energy of the
orbital (010) reflection of the x=0.25 sample near the Mn white
line at 6.547 keV. These data were taken at APS beamline 9ID with
an energy resolution of ~ 1.5 eV. Inset: Close-up of the pre-edge
feature at 6.535 keV.}

\figure{
Polarization-resolved scans of intensity plotted versus incident photon
energy of the orbital (0,1.5,0) and (0,2.5,0) reflections of the x=0.4
sample near the Mn K-absorption edge.}

\figure{ Intensity plotted versus incident photon energy of the
charge (0,3,0) and orbital (0,2.5,0) reflections of the x=0.4
sample for three different values of the azimuthal angle.  The
feature at E=6.62 keV and $\psi$=0 in the charge order scattering
is attributed to multiple scattering.}

\figure{
Azimuthal dependence of the charge and orbital ordering intensities both
on and off the Mn K-edge resonance, as obtained for the x=0.4 sample.}

\figure{
Polarization-resolved scans of the intensity plotted versus incident
photon energy of the charge (0,1,0) and (0,3,0) reflections of the x=0.4
sample near the Mn K-absorption edge.}

\figure{
Intensity plotted versus incident photon energy of the charge (0,1,0)
reflection of the x=0.5 sample, near the Mn K-absorption edge.}

\figure{ (a) Real and imaginary parts of a generic scattering
factor for Mn$^{3+}$ plotted near the K edge. Im(f$_3^+$) is
shifted by 2 eV to obtain the corresponding scattering factor for
Mn$^{4+}$.  (b) Intensity plotted versus incident photon energy
including the Jahn-Teller distortion of the oxygen octahedra and
using the scattering factors shown in (a). (c) Same as in (b), but 
now including a longitudinal Mn displacement.}

\figure{
Temperature dependence of the scattering at the orbital (0,3,0)
reflection of the x=0.25 sample.  Open circles represent data taken on
X22C, open squares taken at beamline X21.  The data sets have been
scaled to agree at room temperature.}

\figure{ Upper: Temperature dependence of the scattering at the
charge (0,3,0) reflection of the x=0.4 sample. Open circles were
obtained on resonance and closed circles off resonance. Lower:
Temperature dependence of the scattering at the orbital (0,1.5,0)
and (0,2.5,0) reflections of the x=0.4 sample. The data at the
(0,1.5,0) reflection (open circles) were obtained on resonance at
the Mn K-edge, while that at the (0,2.5,0) reflection (closed
circles) were obtained off-resonance.  Note that resonant
scattering at the (0,1.5,0) reflection is dominated by the orbital
$\sigma \rightarrow \pi$ contribution in Fig. 7, and not that of
the lattice modulation.}

\figure{
Upper: Temperature dependence of the scattering at the orbital (0,1.5,0)
reflection of the x=0.5 sample. Circles and squares were obtained with a
Ge(111) analyzer, triangles with a Cu(220) analyzer. Lower: Temperature
dependence of the scattering at the charge (0,1,0) reflection obtained
with a Ge(111) analyzer.}

\figure{ Upper: Longitudinal scans of the Bragg (0,2,0), the
charge (0,1,0), and the orbital (0,2.5,0) reflections of the x=0.4
sample at T=8 K. Lower: The same for the x=0.5 sample. Data have
been normalized to the same peak intensity to facilitate
comparison.}

\figure{ Temperature dependence of the half-widths-at-half maximum
of the orbital (0,1,0) reflection of the x=0.25 (diamonds), x=0.4
(0,1.5,0) reflection (squares) and x=0.5 reflection (circles)
samples. Note:  These widths represent the raw data, i.e., without
corrections for resolution effects.}

\figure{
Temperature dependence of the b-axis lattice constants of the x=0.4
(open) and x=0.5 (closed) samples.}

\figure{
Temperature dependence of the orbital and charge wavevectors measured in
all three samples. Crosses give the results of electron diffraction
studies of a sample with x=0.5 taken from reference \cite{Chen99}.}

\figure{
(a) Temperature dependence of the peak intensities of the (0,3,0) charge
order peak (closed circles) and the (0,2.5,0) orbital order peak (open
circles) of the x=0.4 sample. (b) Temperature dependence of the
half-widths-at-half-maximum.}

\figure{
Temperature dependence of the undeconvolved half-widths-at-half-maximum
of the scattering at the orbital (0,2.5,0) and charge (0,1,0)
reflections of the x=0.5 sample.}

\figure{
Magnetic field dependence of the charge and orbital ordering intensities
of the x=0.4 sample obtained at 30 K. Inset: Charge and orbital order
superlattice reflections at T=198 K and 11 T. }

\end{document}